\documentclass[preprint2]{aastex6}
\usepackage{CJK}

\def\orb{{\rm orb}}
\def\rp{R_{\rm p}}
\def\giant{{\rm giant}}

\slugcomment{To Be Submitted To AAS Journals}
\shorttitle{Small Planet-Metallicity Correlation}
\shortauthors{Zhu, Wang \& Huang}

\begin{document}
\begin{CJK*}{UTF8}{gbsn}

\title{Dependence of Small Planet Frequency on Stellar Metallicity Hidden by Their Prevalence}

\author{Wei~Zhu~(祝伟)\altaffilmark{1,2}, Ji~Wang~(王吉)\altaffilmark{3}, and Chelsea~Huang~(黄煦)\altaffilmark{2,4}}

\affil{
$^1$ Department of Astronomy, Ohio State University, 140 W. 18th Ave., Columbus, OH  43210, USA; weizhu@astronomy.ohio-state.edu \\
$^2$ Dunlap Institute, University of Toronto, Toronto, ON, M5S3H4, Canada \\
$^3$ California Institute of Technology, 1200 East California Boulevard, Pasadena, CA 91101, USA \\
$^4$ Centre for Planetary Sciences, University of Toronto at Scarborough, 1265 Military Trail, Toronto, ON, M1C1A4, Canada}

\begin{abstract}
    The dependence of gas giant planet occurrence rate on stellar metallicity has been firmly established. We extend this so-called planet-metallicity correlation to broader ranges of metallicities and planet masses/radii. In particular, we assume that the planet-metallicity correlation is a power law below some critical saturation threshold, and that the probability of hosting at least one planet is unity for stars with metallicity above the threshold. We then are able to explain the discrepancy between the tentative detection and null detection in previous studies regarding the planet-metallicity correlation for small planets. In particular, we find that the null detection of this correlation can be attributed to the combination of high planet occurrence rate and low detection efficiency. Therefore, a planet-metallicity correlation for small planets cannot be ruled out. We propose that stars with metallicities lower than the Solar value are better targets for testing the planet-metallicity correlation for small planets.
\end{abstract}

\keywords{methods: statistical --- stars: abundances --- planetary systems --- planets and satellites: fundamental parameters}

\section{Introduction} \label{sec:introduction}

Our understanding of planet formation benefits from the discoveries of dependences of planet occurrence rate on host star properties. One such correlation that has been widely studied is that between planet occurrence rate and host star metallicity. Based on a sample of four systems, \citet{Gonzalez:1997} first noticed that giant planets were more often found around metal-rich stars, although he considered this correlation as the signature of self-pollution during the planet formation process following the description of \citet{Lin:1996}. Follow-up studies with larger samples confirmed the existence of such a correlation, but attributed this planet-metallicity correlation to be ``primordial'' rather than due to planet pollution \citep{Santos:2000,Pinsonneault:2001,Santos:2001,Santos:2003,Santos:2004,FischerValenti:2005,UdrySantos:2007,Sousa:2008,Johnson:2010,Sousa:2011,Mayor:2011,Santerne:2016}. In particular, \citet{FischerValenti:2005} applied a uniform spectroscopic analysis technique to more than one thousand FGK stars with and without planet detections on the Keck, Lick, and Anglo-Australian Telescope (AAT) planet search surveys, and concluded that stars with exoplanets bear no accretion signature and thus are born in higher metallicity molecular clouds. In addition, they found a power-law fit to the data, suggesting that the formation probability for gas giant planets is proportional to the square of the number of metal atoms. The presence of the planet-metallicity relation in the giant planet regime supports the core accretion model \citep[e.g.,][]{IdaLin:2004,Mordasini:2009} rather than the disk instability model \citep[e.g.,][]{Boss:1997,Boss:2002} as the preferred mechanism for giant planet formation within a few AU separation from the star \citep{Johnson:2010}.

Unlike the well-established planet-metallicity correlation for giant planets, it is still unclear whether smaller planets (planet radius $R_{\rm p}\lesssim4R_\oplus$, or planet mass $M_{\rm p}\lesssim M_{\rm Neptune}$), especially terrestrial planets \citep[$R_{\rm p}\lesssim1.6R_\oplus$,][]{Rogers:2015}, also follow a planet-metallicity correlation. Early studies based on limited numbers of low-mass planets detected in radial velocity surveys suggest that the giant planet-metallicity correlation does not extend quantitatively to the small planet regime \citep{Udry:2006,Sousa:2008,Sousa:2011,Mayor:2011}.

A better constraint on the small planet-metallicity correlation requires a much larger number of small planet detections. This became possible only recently thanks to the \emph{Kepler} Space Telescope \citep{Borucki:2010}. Thousands of small planet candidates have been detected by \emph{Kepler} \citep{Borucki:2011a,Borucki:2011b,Batalha:2013,Burke:2014,Rowe:2015,Mullaly:2015,Coughlin:2016}, and the majority of them are either confirmed or believed to be bona fide planets \citep[e.g.,][]{Fressin:2013,Lissauer:2012,Morton:2016}. Several studies have been conducted to investigate the small planet-metallicity correlation based on the \emph{Kepler} planet catalog. \citet{Buchhave:2012} obtained high-resolution spectra for a sample of 152 stars hosting 226 \emph{Kepler} planet candidates, including 175 with radii smaller than $4R_\oplus$. By comparing the cumulative metallicity distributions of hosts of planets with $\rp>4R_\oplus$ and $\rp<4R_\oplus$, they found that the observed frequency of giant planets as a function of host star metallicity requires a much steeper relation for giant planets than for smaller planets. They also found that small planets could form around stars with a wide range of metallicities. However, because \citet{Buchhave:2012} did not have a reference sample of stars without any transiting planets (SNTP), they were unable to tell whether the hosts of small planets were preferentially metal-rich or not.

\citet{WangFischer:2015} and \citet{BuchhaveLatham:2015} constructed the reference stellar samples in two different ways, used different statistical methods, and reached different conclusions. These two studies used the same sample of stars with transiting planets (STP). This STP sample consists of 405 \emph{Kepler} stars orbited by 600 planet candidates, and the parameters of these host stars are measured spectroscopically \citep{Buchhave:2012,Buchhave:2014}. 
\citet{WangFischer:2015} then constructed a reference sample of Solar-like Kepler stars without detected transiting planets (SNTP). To do so, they adopted the stellar parameters of stars in this sample as determined photometrically by \citet{Brown:2011}, but corrected these parameters for the well-known systematic offsets between photometrically- and spectroscopically-determined stellar parameters. This allowed them to put the parameters of stars in the STP and SNTP samples on a common footing. They divided both STP and SNTP samples into two groups, a metal-poor group with [Fe/H]$<-0.05$ and a metal-rich group with [Fe/H]$>0.05$, and studied the planet occurrence rates in both groups. They found that the occurrence rate of planets with $\rp<4R_\oplus$ in the metal-rich group is about twice as high as that in the metal-poor group. In \citet{BuchhaveLatham:2015}, the reference sample consists of 88 dwarf stars from the asteroseismic sample \citep{Chaplin:2014}, which have stellar parameters spectroscopically determined in the same way as those in the STP sample. By comparing the overall metallicity distributions via the two-sample Kolmogorov-Smirnov (KS) test, \citet{BuchhaveLatham:2015} reported a null detection and 3-$\sigma$ detection of planet-metallicity correlation for planets with radii in the range $R_\oplus<\rp<1.7R_\oplus$ and $1.7R_\oplus<\rp<4R_\oplus$, respectively.

The reliabilities of both studies are undermined because of various systematic effects. The correction to photometric metallicities in \citet{WangFischer:2015} may not be as clean as expected, as argued in \citet{BuchhaveLatham:2015}. On the other hand, the statistical method used in \citet{BuchhaveLatham:2015}, by comparing the cumulative distributions of metallicities of STP and SNTP samples, is less effective, considering that the occurrence rate of small planets is fairly high, and their detection significance was probably inflated as stars hosting multiple planets were counted multiple times. Furthermore, \citet{BuchhaveLatham:2015} selected their SNTP sample from the asteroseismic star sample, which presumably would have stellar properties different from the STP sample because . In fact, the 88 dwarf stars used in the SNTP sample of \citet{BuchhaveLatham:2015} have a mass distribution that is different from that in their STP sample.
\footnote{Between their SNTP$_{\rm dwarf}$ sample and STP sample with small planets ($R_\oplus<\rp<4R_\oplus$), we find the two-sample KS test $p$ value to be $<0.001$ ($>3\sigma$) for the mass distribution. In contrast, the two-sample KS test $p$ value for the metallicity distribution is 0.026 (2.2$\sigma$).}
This systematic bias undermines their claim of the planet-metallicity correlation for planets with $1.7R_\oplus<\rp<4R_\oplus$. In addition, the division at $1.7R_\oplus$ was considered arbitrary according to \citet{Schlaufman:2015}.

Because the formation of primarily rocky planets requires the presence of metals, one would expect that the planet-metallicity correlation should, at least qualitatively, be present for planets with broader ranges of planet mass/radius. If such a correlation exists, we are interested in knowing what kind of form it may take, and furthermore in understanding why it has not been detected in the small planet regime especially in \citet{BuchhaveLatham:2015}. We propose a generalized form to describe the planet-metallicity correlation in Section~\ref{sec:form}, which is extended from the form for the giant planet population. We find that, within this framework, the null detection of the small planet-metallicity correlation can be explained by the combination of two facts: the high occurrence rate and the low detection efficiency of such small/low-mass planets. We use a simple model to demonstrate this point in Section~\ref{sec:method}. We discuss our result in Section~\ref{sec:discussion}.

%%%%%%%%%%%%%%%%%%%%%%%%%%%%%%%%%%%%%%%%%%%%%%%%%%%%%%%%%%%%
\section{The Form of Planet-Metallicity Correlation} \label{sec:form}

The fraction of stars with at least one giant planet in the specified planet parameter space
\footnote{Here ``specified planet parameter space'' can be understood as the region where the planet detection is complete or nearly complete.}
, as a function of metallicity abundance $Z$, is described by \citep{FischerValenti:2005}
\begin{equation} \label{eqn:fv05}
    f(Z) = \mathcal{A} Z^\gamma\ .
\end{equation}
This correlation is valid within a given metallicity range. For example, \citet{FischerValenti:2005} specified a range from $-0.5$ to $0.5$ for their metallicity indicator [Fe/H], and found $\mathcal{A}=0.03$ and $\gamma=2.0$ for a sample of 1040 FGK-type main-sequence stars. 

We now extend the above correlation to a broader range of metallicities. The power-law form of Equation~(\ref{eqn:fv05}) must be broken in order to reconcile the fact that $f(Z)$ as the fraction of stars with planets cannot exceed unity. Therefore, we introduce the following form
\begin{equation} \label{eqn:form}
    f(Z) = \left\{ 
    \begin{array}{cl}
        \left(\frac{Z}{Z_0}\right)^\gamma & {,~(Z<Z_0)} \cr
        1 & {,~(Z \ge Z_0)}
    \end{array}\right.\ .
\end{equation}
The normalization factor $\mathcal{A}$ in Equation~(\ref{eqn:fv05}) is replaced by the saturation point $Z_0$. With this revised form, stars more metal-rich than the \emph{saturation metallicity} $Z_0$ will definitely have at least one planet in the parameter space where this correlation stands. This saturation metallicity is related to the normalization $\mathcal{A}$ in Equation~(\ref{eqn:fv05}) by $Z_0 = \mathcal{A}^{-1/\gamma}$. 

Of course, there would be no saturation if $f(Z)$ in Equation~(\ref{eqn:fv05}) were interpreted as the average number of planets per star (i.e., planet occurrence rate), which was also widely used in various studies \citep[e.g.,][]{Fressin:2013,DongZhu:2013,Petigura:2013,Burke:2015}. However, in deriving the average number of planets per star, the formation and the detection of each planet in a multiple-planet system are assumed to be independent events \citep{Youdin:2011}. This assumption is probably valid when the planet distribution as functions of planetary properties, such as planet mass/radius and orbital period, is studied, but is likely inappropriate if the dependence on stellar properties, such as stellar mass and metallicity, is concerned \citep[e.g.,][]{Cumming:2008,Johnson:2010}. Therefore, the fraction of stars with at least one planet is a more reasonable interpretation of $f(Z)$, with which the introduction of saturation is inevitable.

The distribution of solid material among the planets in any particular system is unlikely to distort the functional form of the planet-metallicity correlation (Equation~\ref{eqn:form}) significantly. First, since $f(Z)$ is the fraction of stars with at least one planet, it is not sensitive to the number of planets in each system. Although a multi-planet system probably undergoes chaotic evolution after its formation, which might largely reshape the whole system, it is very unlikely that any of the chaotic processes could remove all the planets from the system \citep[e.g.,][]{RasioFord:1996,Nagasawa:2008,PuWu:2015}. As long as there is one planet surviving throughout the evolutionary stage, the system is still counted as a planetary system and the fraction of stars with planet(s) remains unaffected.

One may worry about the metal distribution between giant ($\rp>4R_\oplus$) and small ($\rp<4R_\oplus$) planets, because various studies have shown that they are two different populations and thus bear different planet-metallicity correlations \citep{Udry:2006,Sousa:2008,Sousa:2011,Buchhave:2012,Buchhave:2014}. This means that there may be different values for $Z_0$ and/or $\gamma$ for giant and small planets. 
The distribution of solid material between the two planet populations might lead to the planet-metallicity correlation deviating from Equation~(\ref{eqn:form}). For example, one system with enough solid material may preferentially form one giant planet rather than two or more small planets. Therefore, the establishment of Equation~(\ref{eqn:form}) seems to assume that the formations of small and giant planets are unrelated. This is a reasonable assumption for two reasons. First, the core accretion process, which dominates the metal distribution among planets, is believed to be a local behavior \citep[e.g.,][]{KokuboIda:2002}. Second, the dynamical evolution of some giant planets (namely, hot Jupiters) would have significant impact on those small planets in the same system \citep[e.g.,][]{Lin:1996,RasioFord:1996}, but the fraction of such systems is fairly low \citep[e.g.,][]{Mayor:2011,Wright:2012}. We therefore assume that the planet-metallicity correlations for small and giant planets are uncorrelated in our model, but will discuss how this assumption can be relieved in Section~\ref{sec:discussion}.

Finally, the saturation metallicity $Z_0$ is related to the integrated fraction of stars with planet(s). For a sample of stars with a metallicity distribution $g(Z)$, the total fraction of stars with planet(s) is then given by
\begin{equation} \label{eqn:eta-z0}
    \eta = \int f(Z) g(Z) dZ\ .
\end{equation}
The above equation provides a relation between the saturation metallicity $Z_0$ and the total fraction of stars with planet(s) $\eta$. We will use this relation to inversely determine $Z_0$ based on our current knowledge of $\eta$.

%%%%%%%%%%%%%%%%%%%%%%%%%
\section{Method} \label{sec:method}

In this section, we demonstrate with a simple but realistic model how the total fraction of stars with planets, together with the low detection rate, affects the detection of the planet-metallicity correlation. For reasons that have been given in the previous section, each of the planetary systems generated in our simulation has only one planet. Thus, in our simple model, the term ``planet occurrence rate'' is mathematically equivalent to ``the fraction of stars with planet(s)''. 

We use $4R_\oplus$ as the division between giant and small planets. This means that we cannot make detailed or very quantitative comparison with \citet{BuchhaveLatham:2015} and \citet{WangFischer:2015}, because both studies only included planets with $\rp<1.7R_\oplus$. The reason why we choose $4R_\oplus$ rather than $1.7R_\oplus$ is multi-folds. First, the planet-metallicity correlation for planets with $1.7R_\oplus<\rp<4R_\oplus$ is not yet reliably detected (see Section~\ref{sec:introduction}). Second, planets with $1.7R_\oplus<\rp<4R_\oplus$ contribute significantly to the total planet population, and thus must be taken into account when we are simulating the whole observation process through forward modeling. Furthermore, if we only include planets with $\rp<1.7R_\oplus$ in the mock detection process, the average detection efficiency would decrease. Therefore, our result based on planets with $\rp<4R_\oplus$ provides an upper limit on the detection significance for the case with $\rp<1.7R_\oplus$.

\subsection{Model Ingredients}

We restrict ourselves to Sun-like (FGK-type dwarf) stars in the \emph{Kepler} field, which are the primary targets of studies on the small planet-metallicity correlation \citep{Buchhave:2012,WangFischer:2015,BuchhaveLatham:2015}. The metallicity distribution of the underlying stellar sample in the \emph{Kepler} field, $g(Z)$, is taken as a log-normal distribution with mean and dispersion in $\log{Z}$ to be $-0.03$ and $0.20$, respectively. These values are chosen based on the massive low-resolution spectra of 12,000 \emph{Kepler} stars \citep{Dong:2014} from the Large Sky Area Multi-Object Fiber Spectroscopic Telescope survey \citep[LAMOST,][]{Zhao:2012,DeCat:2015}. These values are also consistent with the measurements of red giant stars at the location of the \emph{Kepler} field from SDSS-III/APOGEE survey \citep{Hayden:2015}. Deviations from the adopted metallicity distribution would have effects on the detection of the planet-metallicity correlation. However, according to Equation~(\ref{eqn:eta-z0}), such effects can be accounted as a modest change of the overall planet occurrence rate $\eta$. For simplicity, we therefore choose a single metallicity distribution $g(Z)$, but consider different values of $\eta$.

We assume that the planet-metallicity correlation (Equation~\ref{eqn:form}) applies to planets with orbital periods from 5 days to 4.4 yr \citep[i.e., the snow line for a Sun-like star,][]{KennedyKenyon:2008}, mostly for the purpose of deriving the nominal planet occurrence rate. We remind that these boundaries must be chosen based on physical considerations rather than observational constraints. In particular, one should not only include planets within $\sim200$ day orbit simply because the \emph{Kepler} planet search is close to complete within such a limit. Here the inner boundary (5 days) is adopted because planets inside such an orbit may have undergone significant atmosphere evaporation, so that may show deviation from their primordial planet-metallicity correlation \citep{OwenWu:2013,Buchhave:2014,Lundkvist:2016}. The outer boundary is adopted because planets inside and outside the snow line might show different orbital distributions and/or dependence of stellar properties \citep{IdaLin:2004,Mordasini:2009}. 
The choice of these boundaries only affects the result marginally. Planets beyond the adopted outer boundary have extremely low probability to be detected. In fact, the transit probability for a planet at $P_\orb=4.4$~yr is already $1.7\times10^{-3}$. Planets inside the adopted inner boundary do have larger probability to be detected, but the occurrence rate declines much more dramatically for shorter orbital periods \citep[e.g.,][]{Fressin:2013,DongZhu:2013}. For example, the transit probability increases by a factor of three as the orbit shrinks from 5-day to 1-day period, but the planet occurrence rate is suppressed by more than an order of magnitude \citep{DongZhu:2013}. Furthermore, the impact of choosing a different period interval can be mostly accounted for as a variation in the overall planet occurrence rate. The latter will be discussed in details later.

We describe our choices for the fraction of stars with giant/small planets as follows. For giant planets, \citet{Santerne:2016} found the occurrence rate to be $\sim5\%$ for planets with $\rp\gtrsim6R_\oplus$ and orbital period $P_\orb$ within 400 days. We take this as a lower limit on the fraction of FGK stars with giant planets ($\rp>4R_\oplus$) up to 4.4 yr orbit. The upper limit is chosen to be $15\%$ based on the result from \citet{Mayor:2011}. Therefore, the fraction of stars with giant planets out to 4.4 yr orbit, $\eta_\giant$, is in the range $5\%-15\%$. For small planets ($\rp<4R_\oplus$), \citet{Fressin:2013,DongZhu:2013} and \citet{Petigura:2013} all found the occurrence rate of such planets within 100-day orbit to be $\sim0.5$. We then use the result from \citet{Fressin:2013} to correct for the difference between the average number of planets per star and the fraction of stars with planets, and find that $\sim38\%$ of Sun-like stars hold small planets with orbital period in the range $5-100$ days. Our knowledge of the occurrence rate of small planets beyond $\sim200$-day orbit is very limited, due to the difficulty in detecting such planets via either transit or radial velocity techniques. Therefore, we extrapolate the above result to 4.4 yr orbital period by assuming a flat distribution in $\log{P_\orb}$ \citep{Fressin:2013,DongZhu:2013,Petigura:2013,Burke:2015}. We find that the nominal value of the fraction of stars with small planets out to 4.4 yr orbit, $\eta_{\rm small}$, is $69\%$. Although this value serves as a reference in our model, our conclusion stands as long as $\eta_{\rm small}>50\%$ (see Section~\ref{sec:model}).

We choose $1.8$ as the nominal value for the power-law index $\gamma$ of the planet-metallicity correlation. This is the value found for the giant planets in the \emph{Kepler} field in \citet{Santerne:2016}. We also consider $\gamma$ values up to 3, which is approximately the highest value reported in literature \citep{Neves:2013}.
\footnote{Although a higher value ($3.8\pm1.2$) for $\gamma$ was reported in \citet{Montet:2014}, it is consistent with our choice of the upper limit within 1$\sigma$.}

%%%%%%%%%%%%%%%%%%%%%%%%%
\subsection{Forward Modeling} \label{sec:model}

\begin{figure*}
\centering
\epsscale{0.8}
\plotone{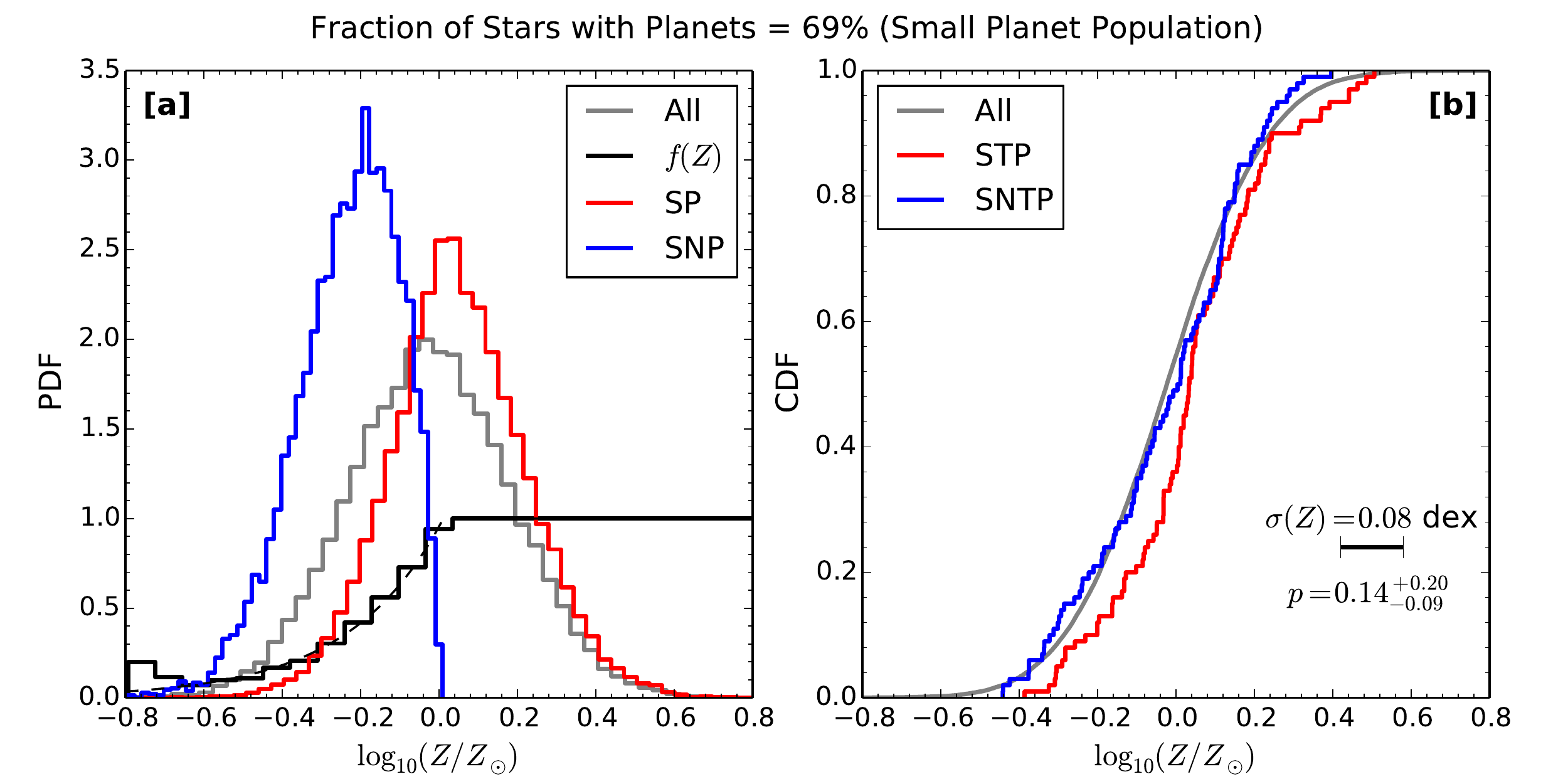}
\plotone{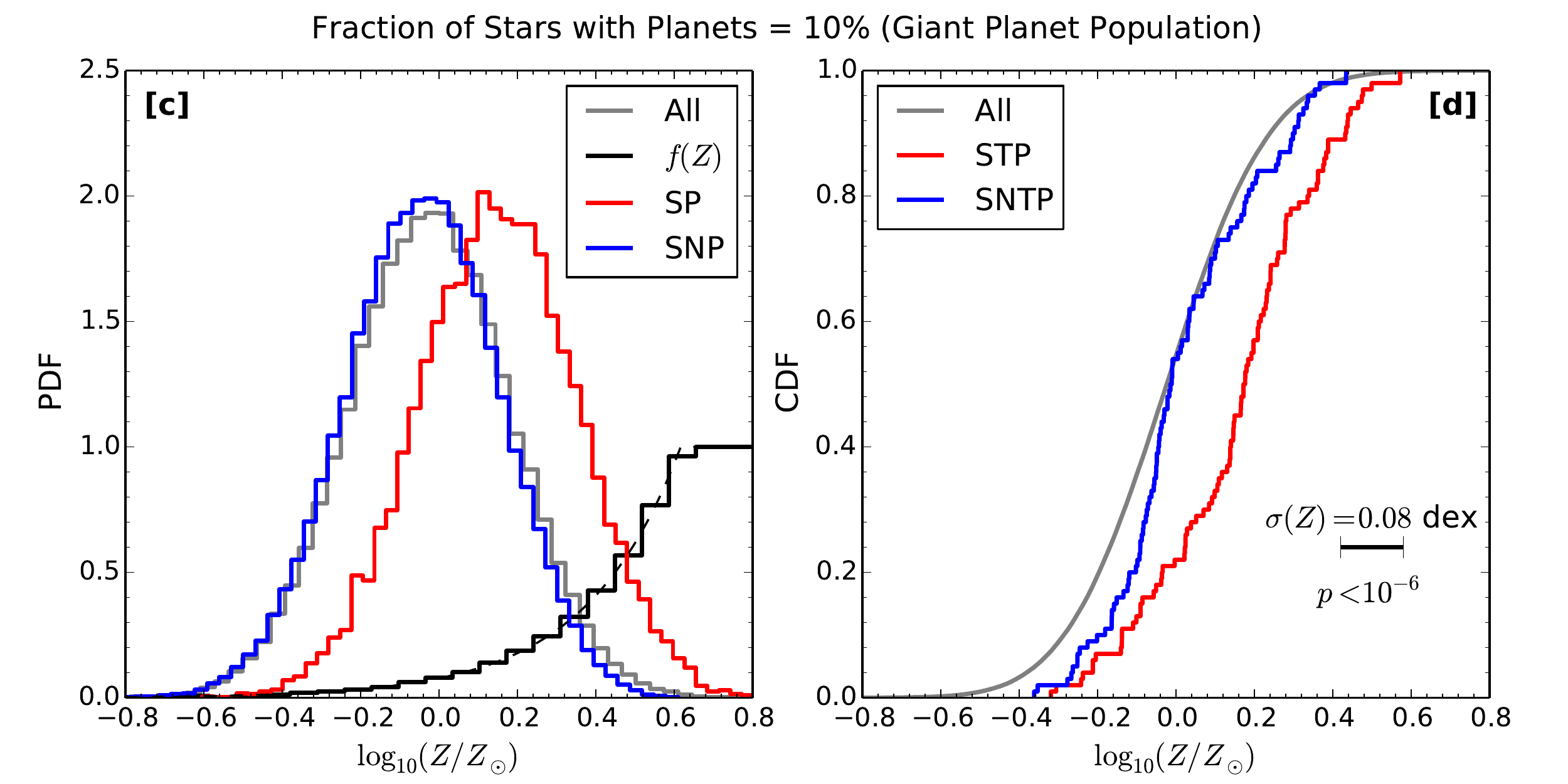}
\caption{Example realizations of our Monte Carlo simulation. We assume that 69\% of stars have planets in the upper panels, and that 10\% of stars have planets in the lower panels. These are typical values for the fractions of stars with small and giant planets, respectively. In each case, the black curve is the underlying planet-metallicity relation $f(Z)$, and the gray histogram is the underlying metallicity distribution $g(Z)$. Based on these, we generate two large samples with and without planets (SP and SNP). When the transit probability is taken into account, these two samples are then re-organized, forming the STP and SNTP samples. We randomly draw 100 stars from each of underlying STP and SNTP samples, and then perform a two-sample KS test. In panels (b) and (d), we show the uncertainty of the metallicity measurement $\sigma(Z)$ and the two-sample KS test $p$ values with 1-$\sigma$ uncertainties. Panels (a) and (c) show the probability distribution functions (PDF), and panels (b) and (d) show the cumulative distribution functions (CDF).
\label{fig:trial}}
\end{figure*}

\begin{figure*}
\centering
\epsscale{0.8}
\plotone{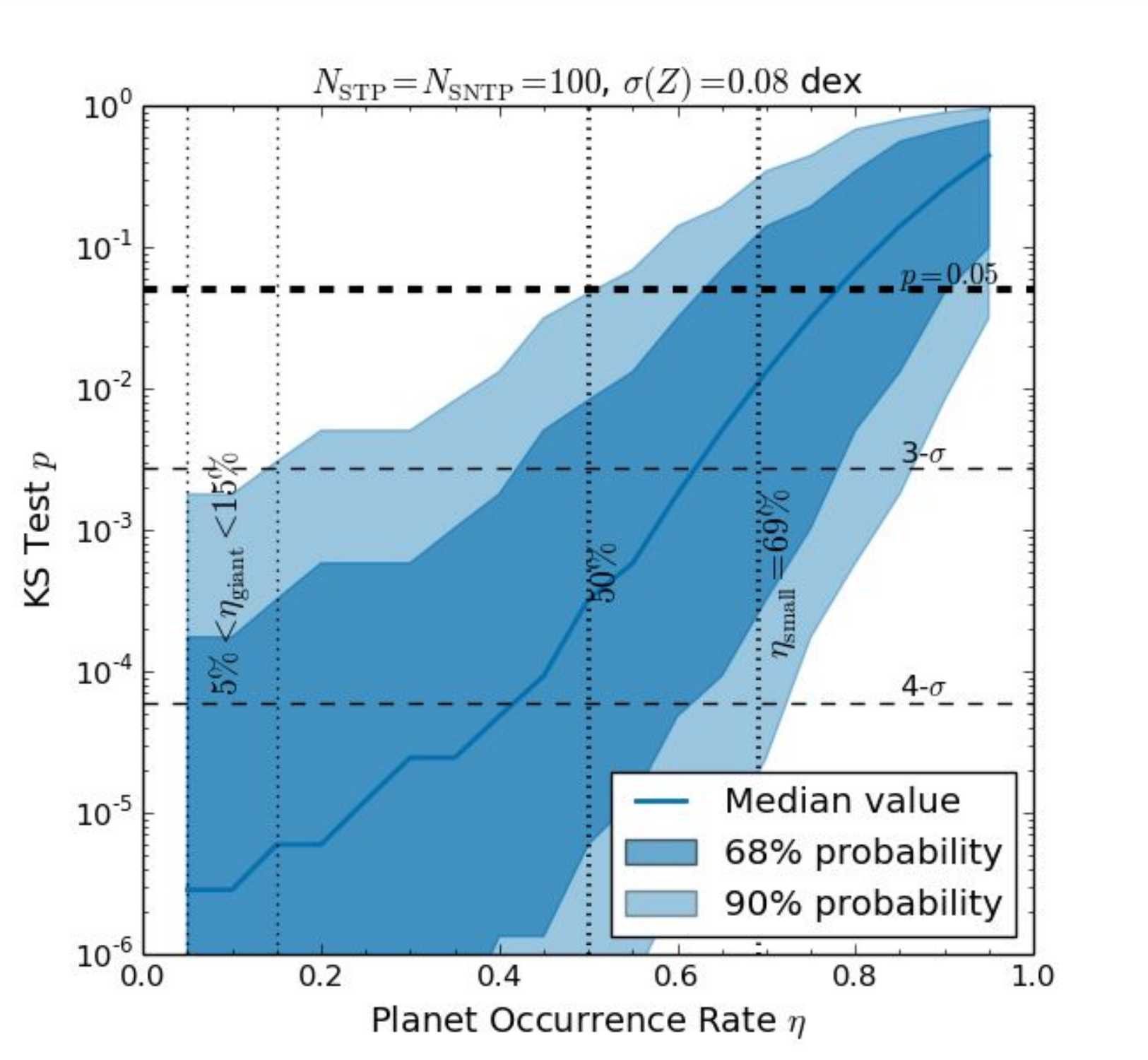}
\caption{The detection significance, quantified by the two-sample KS test $p$ value, as a function of the overall planet occurrence rate $\eta$, for the given sample sizes $N_{\rm STP}=N_{\rm SNTP}=100$ and measurement precision $\sigma(Z)=0.08$ dex. The blue line indicates the median value of $p$ value, and the shades regions enclose $68\%$ and $90\%$ probabilities. The nominal planet occurrence rate for small planets and range for giant planets are also indicated with vertical dotted lines. The input planet-metallicity correlation is considered as being detected if the KS test $p<0.05$.
\label{fig:pval-eta}}
\end{figure*}

With the planet-metallicity correlation described by Equation~(\ref{eqn:form}), we run the following simulation to construct synthetic stellar samples for the detection of the planet-metallicity correlation. 

We first randomly draw metallicities for a significantly large number of stars from the given metallicity distribution $g(Z)$. These stars are then assigned with planets according to $f(Z)$, which quantifies the probability to host a planetary system for a given metallicity. After this step, the original stellar sample is now divided into two: stars with planets (SP) and stars without planets (SNP). Each system in the SP sample is then randomly assigned with following orbital parameters: an inclination $i$ that is drawn from a uniform $\cos{i}$ distribution, and an orbital period $P_{\rm orb}$ that is drawn from a flat distribution in $\log{P_\orb}$ \citep{Opik:1924} between 5 days to 4.4 yrs \citep{Fressin:2013,DongZhu:2013,Petigura:2013,Burke:2015}. Given a host star with mass $M_\odot$ and radius $R_\odot$, if the inclination $i$ and the orbital period $P_\orb$ allow for planet transit, this system falls into the underlying sample of stars with transiting planets (STP$_{\rm all}$). Systems in the SP sample that do not meet the above condition, and all systems in the SNP sample, form the underlying sample of stars without transiting planets (SNTP$_{\rm all}$). In the end, we randomly draw $N_{\rm STP}$ and $N_{\rm SNTP}$ systems from the STP$_{\rm all}$ and SNTP$_{\rm all}$ samples, respectively, to generate the synthetic STP and SNTP samples for the mock detection.

We simulate the detection process following the method that was used in \citet{BuchhaveLatham:2015}. We compare the cumulative distributions of metallicities of the two samples (STP and SNTP), and quantify the significance of the difference by the two-sample Kolmogorov-Smirnov (KS) test $p$ value. To account for the fractional uncertainty in the metallicity measurement $\sigma(Z)$, we conduct 200 realizations: each time the metallicities are randomly drawn from log-normal distributions with standard deviation $\sigma(Z)$ around the nominal values.

For our fiducial run, we choose $N_{\rm STP}=N_{\rm SNTP}=100$, and $\sigma(Z)=0.08$ dex, all of which are similar to those used in \citet{BuchhaveLatham:2015}.
\footnote{\citet{BuchhaveLatham:2015} had 259 and 77 stars in their STP samples with small and giant planets, respectively, and 88 FGK-type dwarf stars in their SNTP sample, so our adopted numbers are not in exact match with these numbers. However, the two-sample KS test is sensitive to the harmonic mean of $N_{\rm STP}$ and $N_{\rm SNTP}$, and thus the deviation of the adopted numbers from the numbers in \citet{BuchhaveLatham:2015} is within 20\%. Furthermore, we notice that the STP and SNTP samples in \citet{BuchhaveLatham:2015} show very different stellar mass distributions. The correction of this selection bias would further undermine their sample sizes.}
    Figure~\ref{fig:trial} demonstrates two fiducial runs with the same power-law index $\gamma=1.8$, one with $\eta=69\%$ (i.e., a typical small planet occurrence rate) and the other with $\eta=10\%$ (i.e., a typical giant planet occurrence rate) in the pre-defined parameter space. For the giant planet population, the two-sample KS test $p$ value is found to be less than $10^{-6}$, suggesting a greater-than $4\sigma$ detection of the input planet-metallicity correlation. By contrast, given a two-sample KS test $p=0.14_{-0.09}^{+0.20}$, the correlation for the small planet population, although with the same $\gamma$ value, cannot be considered as a detection.

    To further quantify how frequent a null detection (KS $p$ value $>0.05$) happens, we generate 1000 sets of synthetic STP and SNTP samples for planet occurrence rate $\eta$ in the range from 5\% to 95\%, and then run the mock detection process to find the two-sample KS test $p$ values. Our results are shown in Figure~\ref{fig:pval-eta}. Here the solid lines show the median $p$ value and the filled regions enclose 68\% and 90\% probabilities. Figure~\ref{fig:pval-eta} confirms our suspicion that the high planet occurrence rate undermines the detection significance of the planet-metallicity correlation. Using the adopted parameters (sample size $N_{\rm STP}=N_{\rm SNTP}=100$, and $\sigma(Z)=0.08$ dex), we find that for the nominal small planet occurrence rate ($69\%$), the probability is $\sim30\%$ that one random set of STP and SNTP samples cannot yield a reliable detection ($p<0.05$), and that in particular, the probability to have a KS $p$ value no less than that (0.14) found in Figure~\ref{fig:trial} is fairly significant (15\%). Furthermore, the chance is greater than 5\% that a random realization cannot yield a detection of the planet-metallicity correlation, as long as more than half of stars hold planets in the specified parameter space. In contrast, the giant planet-metallicity correlation can almost always be reliably detected, and the example shown in the lower panels of Figure~\ref{fig:trial} is very typical (with $\sim$30\% chance).

As an extension, we also consider whether our model explains the null detection of the small planet-metallicity correlation in the radial velocity observations. We keep using the labels STP and SNTP here simply for convenience. We reduce $N_{\rm STP}$ and $N_{\rm SNTP}$ to 23 and 822, in order to match the numbers in \citet{Mayor:2011}. After assigning orbital parameters to the planets in the SP sample, we also randomly assign planet radii, which are drawn from a flat $\log{R_{\rm p}}$ distribution between $R_\oplus$ and $4R_\oplus$, and then estimate the planet masses using the planet mass-radius relation from \citet{WeissMarcy:2014}. The detection criterion is adopted such that the stellar radial velocity semi-amplitude $K>1~$m~s$^{-1}$. Our revised model shows that with current RV sample sizes and precision, a small planet-metallicity correlation with $\gamma=1.8$ remains undetectable even for very low occurrence rate.

Therefore, we are able to explain the null detection of the planet-metallicity correlation in the small planet regime by combining the high occurrence rate and low detection probability (low transit probability or inadequate radial velocity precision) of such planets. A universal planet-metallicity correlation, in the form of Equation~(\ref{eqn:form}), cannot be excluded based on current sample sizes, detection efficiency and data quality.

Our model can also reproduce the tentative detection in \citet{WangFischer:2015} on the occurrence rate enhancement due to stellar metallicity. Using the planet-metallicity correlation $f(Z)$ and the stellar metallicity distribution $g(Z)$ (labelled as ``All'') in Figure~\ref{fig:trial}, we compute the relative planet occurrence rate of metal-rich to metal-poor stars similar to that in \citet{WangFischer:2015}, $(\int_{+0.05}^{+\infty} f(Z) g(Z) d\log{Z})/(\int_{-\infty}^{-0.05} f(Z) g(Z) d\log{Z})$. For planets with sizes below $4R_\oplus$, we find the enhancement to be $2.2$. The enhancement remains the same if planets with sizes below $1.7R_\oplus$ are concerned, because neither $f(Z)$ nor $g(Z)$ depends on the planet size. This value is in good agreement with the number reported in \citet{WangFischer:2015}.

\subsection{Toward a Detection of Small Planet-Metallicity Correlation}

As shown above, the two-sample KS test on the overall metallicity distribution becomes less sensitive to the planet-metallicity correlation once the planet occurrence rate is relatively high. However, given that this statistical approach is simple \citep[compared to the method of][]{WangFischer:2015} and model-independent (compared to the forward modeling method), there is still potential usage of it.

\begin{figure}
\epsscale{1.3}
\plotone{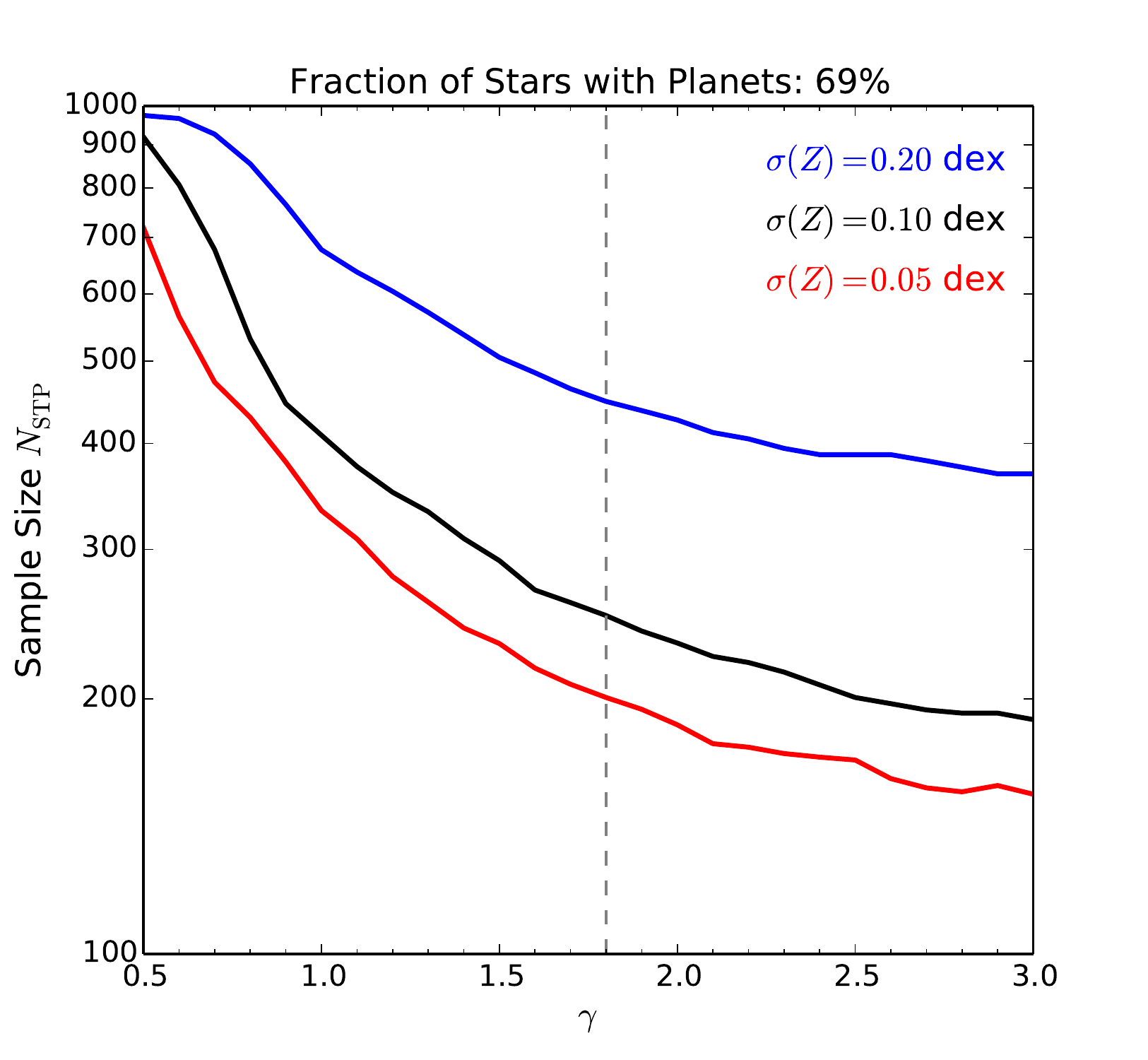}
\caption{For the nominal small planet occurrence rate $\eta=69\%$ and three choices of the metallicity measurement precision, the critical sample size $N_{\rm STP}$ ($=N_{\rm SNTP}$) required to confidently detect the planet-metallicity correlation with a power-law index $\gamma$. Here the term ``confidently'' means that a random set of STP and SNTP samples has greater than $95\%$ probability to detect the input correlation at greater than $2$ sigma confidence level.
\label{fig:sample-sizes}}
\end{figure}

We now investigate that, with this two-sample KS test method, how large the STP and SNTP samples should be in order to detect the planet-metallicity correlation for small planets, assuming there is one. Again by taking advantage of the fact that the two-sample KS test is sensitive to the harmonic mean, we assume both STP and SNTP samples have the same number of stars, in order to reduce the degree of freedoms of our model. Given the nominal small planet occurrence rate $\eta=69\%$, we search for the threshold sample size $N_{\rm STP}$, with which a random realization has a 95\% probability to yield reliable ($p<0.05$) detection of the correlation. We consider three values for the uncertainty of individual metallicity measurement: 5\%, 10\%, and 20\%. Our result is shown in Figure~\ref{fig:sample-sizes}. This result suggests that with current sample size ($N_{\rm STP}=N_{\rm SNTP}=100$) and metallicity precision (0.08 dex), a value for $\gamma$ up to 3 cannot be confidently excluded. It also indicates that, with a precision of 5\% metallicity measurement, if the small planet-metallicity correlation also has a power-law index $\gamma=1.8$, a sample size $N_{\rm STP}$ ($=N_{\rm SNTP}$) nearly twice as large as it is now is required in order to reliably detect the correlation. We note that the STP sample in \citet{BuchhaveLatham:2015} has already accumulated 251 small planet hosts, but a rigorously selected SNTP sample has not reached an equivalent size. 

The \citet{WangFischer:2015} approach, as a simplified version of the method used in \citet{FischerValenti:2005}, requires metallicity measurements of an even larger sample of stars in the SNTP sample, or a reliable calibration of between the photometric and spectroscopic metallicities. However, this latter approach has the advantage of quantifying the parameter $\gamma$, as has been demonstrated by various studies on the giant planet-metallicity correlation \citep[e.g.,][]{FischerValenti:2005,UdrySantos:2007}

In addition to simply acquiring metallicity measurements for more stars, another potential improvement would be utilizing stars with metallicities in the range where the correlation has large dynamic change. As shown in the upper left panel of Figure~\ref{fig:trial}, our model suggests that $f(Z)$ for small planets saturates around Solar metallicity, predicting that stars with sub-Solar metallicities are perhaps better targets for detecting $\gamma$.

\section{Discussion} \label{sec:discussion}

We start from the naive expectation that the planet-metallicity correlation should at least qualitatively hold for planets smaller or less massive than Neptune. We then derive the general form of this planet-metallicity correlation (Equation~\ref{eqn:form}), by extending the functional form that has been widely used for giant planets to a broader range of metallicities. The saturation metallicity $Z_0$ is therefore introduced for the purpose to reconcile the monotonic increasing behavior of $f(Z)$ as a power law of metallicity $Z$ and the fact that $f(Z)$ as the fraction of stars with planets should not exceed unity. 

With this general form, we demonstrate with a simple but realistic model that the null detection of the planet-metallicity correlation in the small planet regime in \citet{BuchhaveLatham:2015} and radial velocity studies \citep[e.g.,][]{Sousa:2011,Mayor:2011} can be explained by 1) more than half of stars host at least one small planet, and 2) the methods that are used to detect such planets are very inefficient. For the transit method, the detection efficiency is limited by the transit probability to typically a few percent. For radial velocity method, the stellar radial velocity semi-amplitude induced by most of the low-mass planets are below the current detection limit. Because of the high planet occurrence rate and low detection efficiency, the sample of stars without any detected planets is in fact contaminated by a significant fraction of stars having (undetectable) planets. Therefore, the difference between the metallicities of planet hosts and reference stars is significantly reduced, even though the input planet-metallicity correlation is strong.

Our model also reproduces the result in \citet{WangFischer:2015}, namely the enhancement on small planet occurrence rate due to the stellar metallicity. Therefore, we are able to reconcile the two observational results \citep{WangFischer:2015,BuchhaveLatham:2015} that were thought contradictory.

Our model is simple but nevertheless realistic in the sense that it captures the main features that are necessary to reproduce observations. First, because the fraction of stars with planets is concerned, we assume that all planetary systems have only one planet in order to avoid the complexity arisen from multiple-planet systems. After all, we do not intend to quantify the difference between the metallicity distribution of stars with giant planets and that with small planets to compare with \citet{Buchhave:2012} and \citet{Buchhave:2014}, because that would require a detailed forward modeling of the multi-planet system formation and/or a better handle on the coupling of the small and giant planet-metallicity correlations. When simulating the planet detection process, we only consider the intrinsic detection limits such as the transit probability and the stellar radial velocity semi-amplitude, and ignore other observational limitations such as the signal-to-noise ratio of transit signals and the duration of RV observations. We did not take into account the uncertainties on other observables except the stellar metallicities either. The inclusion of all these observational products could only further reduce the detection significance of the planet-metallicity correlation.

In order to detect the planet-metallicity correlation for small planets, if there is one, metallicity measurements of more stars are needed. For transit method, which is currently the most efficient technique to detect such small planets, it requires metallicity measurements of a larger sample of stars without any detected planets. Given the form of the correlation, we also suggest that stars with sub-Solar metallicities would be better targets for detecting the small planet-metallicity correlation.

We find that, given current observational constraints, it is still possible that the planet-metallicity correlation for small planets has the same power-law index $\gamma$ as that for giant planets. As discussed in Section~\ref{sec:form}, the planet-metallicity correlations for small and giant planets are in general assumed to be decoupled. This assumption is no longer necessary with the two correlations sharing the same $\gamma$, as long as $f(Z)$ in Equation~(\ref{eqn:form}) is interpreted as the fraction of stars with planets larger than a given size/mass (rather than planets of either giant or small population).

The theoretical implication is profound if the small planet-metallicity correlation shares the same power-law index as the giant planet-metallicity correlation. The formation of planets within a few AU separation from their hosts is believed to be through accretion of either km-sized planetesimals \citep[e.g.,][]{KokuboIda:2002,IdaLin:2004,Raymond:2006,Mordasini:2009} and/or centimeter- to meter-sized pebbles \citep[e.g.,][]{JohansenLacerda:2010,OrmelKlahr:2010}. Such a formation mechanism intrinsically requires $\gamma\sim2$ regardless of the size of the planet, because the overall particle collision rate is proportional to the square of the number of particles. The efficiency of the accretion process relies on the total amount of solid material in the disk. For giant planets ($\rp>4R_\oplus$), the accretion process needs to be efficient as to form a massive rocky core ($5-10~M_\oplus$) before the gas is depleted in the protoplanetary disk. This massive core could then initiate the run-away gas accretion and eventually grows into a gas giant. Therefore, the more massive the disk is the more likely a giant planet can be formed \citep{FischerValenti:2005}, and the relatively high saturation metallicity is justified. Such an efficient accretion process is not required for the formation of small planets ($\rp<4R_\oplus$), and thus a much lower saturation metallicity is sufficient.

\acknowledgements
We would like to thank Yanqing Wu, Norm Murray, and especially Scott Gaudi for stimulating discussions. 
We thank Andy Gould and Scott Gaudi for comments on the manuscript.
WZ would like to thank the Dunlap Institute for Astronomy and Astrophysics at the University of Toronto for its hospitality. 
The Dunlap Institute is funded through an endowment established by the David Dunlap family and the University of Toronto.
We offer our thanks and praise to the extraordinary scientists, engineers and individuals who have made the \emph{Kepler} Mission possible. 
WZ acknowledges the support from NSF grant AST-1516842.
JW acknowledges the support from JPL RSA No.1533314.

\end{CJK*}
\end{document}